
%
%

\input phyzzx
\tolerance=1000
\sequentialequations
\def\rl{\rightline}

\def\t1{{\tilde 1}}

\def\AEF{A.E. Faraggi}
\def\DVN{D.V. Nanopoulos}

\def\SSM{supersymmetric standard model}
\def\NPB#1#2#3{Nucl. Phys. B {\bf#1} (19#2) #3}
\def\PLB#1#2#3{Phys. Lett. B {\bf#1} (19#2) #3}

\def\MODA#1#2#3{Mod. Phys. Lett. A {\bf#1} (19#2) #3}

\REF\GCU{V. Kaplunovsky, \NPB{307}{88}{145}; I. Antoniadis, J. Ellis,
R. Lacaze and \DVN, CERN--TH 6136/91 (1991).}
\REF\EKN{J. Ellis, S. Kelley and \DVN, CERN--TH.6140/91;
P. Langacker, University of Pennsylvania preprint UPR--0435T (1990);
U. Amaldi, W. de Boer and F. F\"urstenau, \PLB{260}{91}{447}.}
\REF\ILR{L.E. Iba{\~n}ez, D. L\"ust and G.G. Ross,
\PLB{272}{91}{251}.}
\REF\PRICE{I. Antoniadis, J. Ellis, S. Kelley,
and \DVN, \PLB{272}{91}{31}; I.Antoniadis, G.K. Leontaris
and N.D. Tracas, CPTH--A112.1291.}
\REF\SSM{L.E. Iba{\~n}ez {\it{et al.}}, Phys.Lett.
{\bf B191}(1987) 282; A. Font {\it{et al.}},
Phys.Lett. {\bf B210} (1988)
101; A. Font {\it{et al.}},
Nucl.Phys. {\bf B331} (1990)
421; D. Bailin, A. Love and S. Thomas,
Phys.Lett.{\bf B194} (1987) 385;
Nucl.Phys.{\bf B298} (1988) 75; J.A. Casas, E.K. Katehou and C. Mu{\~n}oz,
Nucl.Phys.{\bf B317} (1989) 171.}
\REF\FNY{\AEF, D.V. Nanopoulos and K. Yuan, \NPB{335}{90}{437}.}
\REF\EU{\AEF,  \PLB{278}{92}{131}.}
\REF\TOP{\AEF, \PLB{274}{92}{47}.}
\REF\SLM{\AEF, \NPB{387}{92}{239}.}
\REF\FIQ{A. Font, L.E. Iba{\~n}ez and F. Quevedo,
 Phys.Lett.{\bf B228} (1989) 79; A.E. Faraggi and D.V. Nanopoulos,
\MODA{6}{91}{61}; \AEF, \PLB{245}{91}{437}.}
\REF\DSW{M. Dine, N. Seiberg and E. Witten, Nucl.Phys.{\bf B289} (1987) 585.}
\REF\FFF{I. Antoniadis and C. Bachas, Nucl.Phys.{\bf B298} (1988)
586; H. Kawai, D.C. Lewellen, and S.H.-H. Tye,
Nucl.Phys.{\bf B288} (1987) 1;
R. Bluhm, L. Dolan, and P. Goddard, Nucl.Phys.{\bf B309} (1988) 330.}
\REF\naturalness{A.E. Faraggi and D.V. Nanopoulos,
Texas A \& M University preprint CTP--TAMU--78, ACT--15;
\AEF, Ph.D thesis, CTP--TAMU--20/91, ACT--31.}
\REF\YUKAWA{\AEF, WIS--91/83/--PH.}
\REF\GHS{A. Giveon, L.J. Hall and U. Sarid, \PLB{271}{91}{138}.}

\singlespace
\rl{WIS--92/17/FEB--PH}
\rl\today
\normalspace
\medskip
\titlestyle{\bf Gauge Coupling Unification in Superstring Derived
Standard--like Models}
\author{Alon E. Faraggi{\footnote*{e--mail address: fhalon@weizmann.bitnet}}}
\medskip
\centerline {Department of physics, Weizmann Institute of Science}
\centerline {Rehovot 76100, Israel}
\titlestyle{ABSTRACT}

I discuss gauge coupling unification
in a class of superstring standard--like models,
which are derived in the free fermionic formulation.
Recent calculations indicate that the
superstring unification scale is at $O(10^{18}GeV)$ while the minimal
supersymmetric standard model is consistent with LEP data if the
unification scale is at $O(10^{16})GeV$.
A generic feature of the superstring standard--like
models is the appearance of extra
color triplets $(D,{\bar D})$, and electroweak doublets $(\ell,{\bar\ell})$,
in  vector--like  representations,  beyond the minimal
supersymmetric standard  model.
I show that gauge coupling unification
at $O(10^{18}GeV)$ in the superstring standard--like
models can be consistent with LEP data. I present an explicit
standard--like model that can realize superstring
gauge coupling unification.

\singlespace
\vskip 0.5cm
\centerline{To appear in Phys. Lett. {\bf B}}
\endpage
\normalspace
\pagenumber 1

\noindent{\bf 1. Introduction}

Superstring theory
is a unique candidate for the unification of
gravity with the gauge interactions. As unified theories
of the gauge interactions,  superstring theories
predict that the gauge coupling unification scale
is at O($10^{18}$ GeV) [\GCU]. On the other hand, recent precision LEP data
for ${\sin^2}{\theta_W}$ and $\alpha_s$, indicates that the unification
scale of the minimal supersymmetric
standard model is at O($10^{16}$ GeV) [\EKN].
Thus, two orders of magnitude
separate the superstring unification scale  and the
successful unification scale of SUSY GUTS.

This problem is one of the challenges facing superstring theory.
It does not seem to be possible to resolve this problem
by invoking uncertainties in the low energy inputs,
uncertainties in the large extrapolation due to higher loop
effects or by inclusion of the effect of Yukawa couplings on
the renormalization group equations.
Several solutions have been contemplated to resolve this problem .
In Ref. [\ILR] the effects of string threshold corrections
were examined in detail. From the results we may conclude that string
threshold corrections are not likely to resolve the problem.
In Ref. [\PRICE] possible extensions of the spectrum of the
minimal supersymmetric standard model were speculated in the context
of superstring GUT models.

An  attractive alternative to superstring models which are based on
intermediate GUT models is to derive the Standard Model directly
from the superstring [\SSM,\FNY,\EU,\TOP,\SLM]. In these models
the gauge couplings of the Standard Model must unify at the
superstring unification scale.
Among the superstring models, there is a unique class of standard-like
models [\EU,\TOP,\SLM] with the following properties:

\parindent=-15pt

1. Three and only three generations of chiral fermions.

2. The gauge group is ${SU(3)_C}\times{SU(2)_L}\times{U(1)_{B-L}}\times
{U(1)_{{T_3}_R}}\times U(1)^n\times{hidden}$. $n$ reduces to one or
zero after application of the Dine--Seiberg--Witten mechanism.
The
$U(1)_{Z^\prime}={1\over2}{U(1)_{B-L}}-{2\over3}{U(1)_{{T_3}_R}}$
combination may be broken at
the Planck scale. If it remains unbroken down to low energies, it results in a
gauged mechanism to suppress proton decay from
dimension four operators [\FIQ].

3. There are enough scalar doublets and singlets
to break the symmetry in a realistic way and to generate realistic fermion
mass hierarchy [\TOP,\SLM].

4. Proton decay from dimension four and dimension five operators is suppressed
due to gauged $U(1)$ symmetries [\SLM].

5. These models explain the top-bottom mass hierarchy. At the trilinear level
of the superpotential, only the top quark gets a non--vanishing mass term.
The mass terms for the bottom quark and for the lighter quarks and leptons are
obtained from non--renormalizable terms. Thus, only the top quark mass is
characterized by the electroweak scale and the masses of the lighter quarks and
leptons are naturally suppressed [\TOP,\YUKAWA].
The top--bottom mass hierarchy is correlated with the
requirement of a supersymmetric vacuum
at the Planck scale [\EU,\TOP,\SLM].

\parindent=15pt

In this paper I address the question of gauge coupling unification in the
superstring derived standard--like models.
A generic feature of these models is the appearance of additional
color triplets, $(D,{\bar D})$, and electroweak doublets $(\ell,{\bar\ell})$,
in  vector--like  representations,  beyond the minimal
standard supersymmetric model.
I show that gauge coupling unification  at
$O(10^{18})GeV$ can be consistent with LEP precision data for $\sin^2\theta_W$
and $\alpha_s$, provided that the additional states are found at the
appropriate scales. I construct a specific standard--like model
which can realize superstring gauge coupling unification.

\bigskip
\noindent{\bf 2. The superstring model}

The superstring standard--like models are derived in the free fermionic
formulation [\FFF].  In this formulation all the degrees
of freedom needed to cancel
the conformal anomaly are represented
in terms of internal free fermions propagating
on the string world--sheet.
Under parallel transport around a non-contractible loop,
the fermionic states pick up a phase.
Specification of the phases for all world--sheet fermions
around all noncontractible loops contribute
to the spin structure of the model.
The possible spin structures are constrained by string consistency requirements
(e.g. modular invariance).
A model is constructed by choosing a set of boundary condition vectors,
which satisfies
the modular invariance constraints.
The basis vectors, $b_k$, span a finite
additive group $\Xi=\sum_k{{n_k}{b_k}}$
where $n_k=0,\cdots,{{N_{z_k}}-1}$.
The physical massless states in the Hilbert space of a given sector
$\alpha\in{\Xi}$, are obtained by acting on the vacuum with
bosonic and fermionic operators and by
applying the generalized GSO projections.

The superstring model is generated by a basis of
eight vectors of boundary conditions
for all the world--sheet fermions. The first five vectors in the
basis consist of the NAHE set, $\{{{\bf 1},S,b_1,b_2,b_3}\}$
[\naturalness,\SLM].
In addition to the first five
vectors, the basis contains three additional vectors.
These vectors and the choice of generalized GSO projection coefficients
are given in table 1, where the notation of Refs. [\YUKAWA, \SLM] is used.
The basis vectors $\alpha$, $\beta$ and $\gamma$ are the basis vectors
of Ref. [\TOP]. The following generalized GSO projection coefficients
are modified
$$c\left(\matrix{1\cr
                                    1\cr}\right)=
-c\left(\matrix{\gamma\cr
                                    1\cr}\right)=-1.\eqno(1)$$
The set {$\{1,S,b_1,b_2,b_3\}$}
gives an $N=1$ supersymmetric,
$SO(10)\times SO(6)^3\times E_8$
gauge group with $3\times 2$ copies of massless chiral fields
$(16+4)+(16+\bar 4)$, two from each of the sectors $b_1,b_2,b_3$.
 The vectors $\alpha,\beta,\gamma$ break the horizontal symmetries to
$U(1)^3\times U(1)^3$, which correspond to the right--moving world--sheet
currents ${\bar\eta}^a_{1\over2}{{\bar\eta}^{a^*}}_{1\over2}$
($a=1,2,3$) and
${{\bar y}_3{\bar y}_6,{\bar y}_1{\bar\omega}_5,
{\bar\omega}_2{\bar\omega}_4}$ respectively
(I define $\zeta^1{\equiv}{1\over{\sqrt2}}(\bar y^3+i\bar y^6),
\zeta^2{\equiv}{1\over\sqrt2}(\bar y^1+i\bar\omega^5),
\zeta^3{\equiv}{1\over\sqrt2}(\bar\omega^2+i\bar\omega^4)$).
The vectors $\alpha,\beta$ break the $SO(10)$ symmetry to $SO(6)\times SO(4)$
and the hidden group from $E_8$ to $SO(16)$. The vector $\gamma$
breaks $SO(6)\times SO(4)\rightarrow
SU(3)_C\times U(1)_C\times SU(2)_L\times U(1)_L${\footnote*{
 $U(1)_C={3\over 2}U(1)_{B-L}$ and
$U(1)_L=2U(1)_{T_{3_R}}$.}}.
and $SO(16)\rightarrow SU(5)_H\times SU(3)_H\times U(1)^2$.
The weak hypercharge is given by
$U(1)_Y={1\over 3} U(1)_C + {1\over 2} U(1)_L$. The orthogonal
combination is given by $U(1)_{Z^\prime}= U(1)_C - U(1)_L$.
The $U(1)$ symmetries in the
hidden sector, $U(1)_7$ and $U(1)_8$,
correspond to the world--sheet currents
${\bar\phi}^1{\bar\phi}^{1^*}-{\bar\phi}^8{\bar\phi}^{8^*}$ and
$-2{\bar\phi}^j{\bar\phi}^{j^*}+{\bar\phi}^1{\bar\phi}^{1^*}
+4{\bar\phi}^2{\bar\phi}^{2^*}+{\bar\phi}^8{\bar\phi}^{8^*}$ respectively,
where summation on $j=5,\cdots,7$ is implied.

The full massless spectrum was derived by using a FORTRAN
program. Here I list only the states which are relevant
for gauge coupling unification. The full massless spectrum will
be presented elsewhere.
The following massless states are produced by the sectors
$b_{1,2,3}$, $S+b_1+b_2+\alpha+\beta$, $O$ and their superpartners in the
observable $[(SU(3)_C,U(1)_C);(SU(2)_L,$ $U(1)_L)]_{_{1,2,3,4,5,6}}$ sector,
where $1,\cdots,6$ denote the charges under the six extra $U(1)$s.

(a) The $b_{1,2,3}$ sectors produce three $SO(10)$ chiral generations.
$G_\alpha=e_{L_\alpha}^c+u_{L_\alpha}^c+N_{L_\alpha}^c+d_{L_\alpha}^c+
Q_\alpha+L_\alpha$ $(\alpha=1,\cdots,3)$ where
$$\eqalignno{{e_L^c}&\equiv [(1,{3\over2});(1,1)];{\hskip .6cm}
{u_L^c}\equiv [({\bar 3},-{1\over2});(1,-1)];{\hskip .2cm}
Q\equiv [(3,{1\over2});(2,0)]{\hskip 2cm}
&(2a,b,c)\cr
{N_L^c}&\equiv [(1,{3\over2});(1,-1)];{\hskip .2cm}
{d_L^c}\equiv [({\bar 3},-{1\over2});(1,1)];{\hskip .6cm}
L\equiv [(1,-{3\over2});(2,0)]{\hskip 2cm}
&(2d,e,f)\cr}$$
of $SU(3)_C\times U(1)_C\times SU(2)_L\times U(1)_L$, with charges under the
six horizontal $U(1)$s. From the sector $b_1$ we obtain
$${\hskip .2cm}  ({e_L^c}+{u_L^c})_{{1\over2},0,0,{1\over2},0,0}+
({d_L^c}+{N_L^c})_{{1\over2},0,0,{{1\over2}},0,0}+
(L)_{{1\over2},0,0,-{1\over2},0,0}+(Q)_{{1\over2},0,0,-{1\over2},0,0},
\eqno(3a)$$ from the sector $b_2$
$${\hskip .2cm} ({e_L^c}+{u_L^c})_{0,{1\over2},0,0,{1\over2},0}+
({N_L^c}+{d_L^c})_{0,{1\over2},0,0,{1\over2},0}+
(L)_{0,{1\over2},0,0,-{1\over2},0}+
(Q)_{0,{1\over2},0,0,-{1\over2},0},\eqno(3b)$$ and from the sector $b_3$
$$ {\hskip .2cm} ({e_L^c}+{u_L^c})_{0,0,{1\over2},0,0,{1\over2}}+
({N_L^c}+{d_L^c})_{0,0,{1\over2},0,0,{1\over2}}+
(L)_{0,0,{1\over2},0,0,-{1\over2}}+(Q)_{0,0,{1\over2},0,0,-{1\over2}}.
\eqno(3c)$$
The vectors $b_1,b_2,b_3$ are the only vectors in the additive group
$\Xi$ which give rise to spinorial $16$ of $SO(10)$.
The fact that there are exactly three
generations in the standard--like models and no mirror
generations is closely related to
the choice of $SU(3)_C\times U(1)_C\times SU(2)_L\times U(1)_L$
as the observable gauge symmetry at the level of the spin structure.

(b) The ${S+b_1+b_2+\alpha+\beta}$ sector gives
$$\eqalignno{h_{45}&\equiv{[(1,0);(2,-1)]}_
{{1\over2},{1\over2},0,0,0,0} {\hskip .5cm}
{h}_{45}^\prime\equiv{[(1,0);(2,-1)]}_
{-{1\over2},-{1\over2},0,0,0,0}&(4a,b)\cr
\Phi_{45}&\equiv{[(1,0);(1,0)]}_
{-{1\over2},-{1\over2},-1,0,0,0}  {\hskip .5cm}
\Phi^{\prime}_{45}\equiv{[(1,0);(1,0)]}_
{-{1\over2},-{1\over2},1,0,0,0}&(4c,d)\cr
\Phi_1&\equiv{[(1,0);(1,0)]}_
{-{1\over2},{1\over2},0,0,0,0} {\hskip .5cm}
\Phi_2\equiv{[(1,0);(1,0)]}_
{-{1\over2},{1\over2},0,0,0,0}&(4e,f)\cr}$$
(and their conjugates ${\bar h}_{45}$, etc.).
The states are obtained by acting on the vacuum
with the fermionic oscillators
${\bar\psi}^{4,5},{\bar\eta}^3,{\bar y}_5,{\bar\omega}_6$,
respectively  (and their complex conjugates for ${\bar h}_{45}$, etc.).

(c) The Neveu-Schwarz $O$ sector gives, in addition to  the graviton,
dilaton, antisymmetric tensor and spin 1 gauge bosons,  the
following scalar representations:
$$\eqalignno{
{h_1}&\equiv{[(1,0);(2,-1)]}_{1,0,0,0,0,0}
{\hskip .5cm}
\Phi_{23}\equiv{[(1,0);(1,0)]}_{0,1,-1,0,0,0}&(5a,b)\cr
{h_2}&\equiv{[(1,0);(2,-1)]}_{0,1,0,0,0,0} {\hskip .5cm}
\Phi_{13}\equiv{[(1,0);(1,0)]}_{1,0,-1,0,0,0}&(5c,d)\cr
{h_3}&\equiv{[(1,0);(2,-1)]}_{0,0,1,0,0,0}{\hskip .5cm}
\Phi_{12}\equiv{[(1,0);(1,0)]}_{1,-1,0,0,0,0}
&(5e,f)\cr}$$
(and their conjugates ${\bar h}_1$ etc.).
Finally, the Neveu--Schwarz sector gives rise to three singlet
states that are neutral under all the U(1) symmetries.
$\xi_{1,2,3}:{\hskip .2cm}{\chi^{12}_{1\over2}{\bar\omega}^3_{1\over2}
{\bar\omega}^6_{1\over2}{\vert 0\rangle}_0},$
 ${\chi^{34}_{{1\over2}}{\bar y}_{1\over2}^5{\bar\omega}_{1\over2}^1
{\vert 0\rangle}_0},$
 $\chi^{56}_{1\over2}{\bar y}_{1\over2}^2{\bar y}_{1\over2}^4
{\vert 0\rangle}_0.$

In addition to these states the model contains additional
color triplets and electroweak doublets in vector--like representations.
These states carry $U(1)$ charges under the
hidden $U(1)$ symmetries.
The sectors in the additive group,
which produce these states are obtained from a combination of
$\gamma$, or $2\gamma$, with some combination
of the other basis vectors. The additional color triplets, and
electroweak doublets, and their
quantum numbers are shown in table 2. The number of additional
states is model dependent. The choice of GSO projection coefficients,
Eq. (1), leads to color triplets from the sectors
$b_1+b_3+\alpha\pm\gamma+(I)$ and $b_2+b_3+\beta\pm\gamma+(I)$.
If the opposite sign is taken, as in Ref. [\TOP],
electroweak doublets are obtained. The electroweak doublets in table
2 exist in some models. However, they are not a generic feature of the
the standard--like models. The model of Ref. [\EU] is one example
in which this kind of doublets do not exist in the massless spectrum.
All models have at least one pair of color triplets and one pair
of electroweak doublets beyond the spectrum of the minimal supersymmetric
standard model.
I emphasize that all the color triplets and electroweak doublets, in table 2,
are present in the massless spectrum of the model of table 1.
These additional color triplets, and
electroweak doublets,
can affect the renormalization group equations
in a way which enables gauge coupling unification at the string level.

The model contains three anomalous $U(1)$ symmetries:
Tr${U_1}=24$, Tr${U_2}=24$, Tr${U_3}=24$.
Of the three anomalous $U(1)$s,  two can be rotated by
an orthogonal transformation. One combination remains anomalous and is
uniquely given by: ${U_A}=k\sum_j [{Tr {U(1)_j}}]U(1)_j$,
where $j$ runs over all the
anomalous $U(1)$s.
For convenience, I take $k={1\over{24}}$. Therefore,
the anomalous combination
is given by:
$$U_A=U_1+U_2+U_3,{\hskip 3cm}TrQ_A=72.\eqno(6a)$$
The two orthogonal combinations are not unique. Different
choices are related by orthogonal transformations. One choice is given by:
$$\eqalignno{{U^\prime}_1&=U_1-U_2{\hskip .5cm},{\hskip .5cm}
{U^\prime}_2=U_1+U_2-2U_3.&(6b,c)\cr}$$

I now turn to the superpotential of the  model.
All the non vanishing trilevel and quartic level
terms in the superpotential of the model were generated by a simple
FORTRAN program. I list here only the relevant terms.
At trilevel the following terms are obtained
$$\eqalignno{W_3&=\{(
{u_{L_1}^c}Q_1{\bar h}_1+{N_{L_1}^c}L_1{\bar h}_1+
{u_{L_2}^c}Q_2{\bar h}_2+{N_{L_2}^c}L_2{\bar h}_2+
{u_{L_3}^c}Q_3{\bar h}_3+{N_{L_3}^c}L_3{\bar h}_3)\cr
&\qquad
+{{h_1}{\bar h}_2{\bar\Phi}_{12}}
+{h_1}{\bar h}_3{\bar\Phi}_{13}
+{h_2}{\bar h}_3{\bar\Phi}_{23}
+{\bar h}_1{h_2}{\Phi_{12}}
+{\bar h}_1{h_3}{\Phi_{13}}
+{\bar h}_2{h_3}{\Phi_{23}}\cr
&\qquad
+\Phi_{23}{\bar\Phi}_{13}{\Phi}_{12}
+{\bar\Phi}_{23}{\Phi}_{13}{\bar\Phi}_{12}
+{\bar\Phi}_{12}({\bar\Phi}_1{\bar\Phi}_1
+{\bar\Phi}_2{\bar\Phi}_2)
+{{\Phi}_{12}}(\Phi_1\Phi_1
+\Phi_2\Phi_2)\cr
&\qquad
+{1\over2}\xi_3(\Phi_{45}{\bar\Phi}_{45}
+h_{45}{\bar h}_{45}+{\Phi_{45}^\prime}{{\bar\Phi}_{45}^\prime}
+h_{45}^\prime{\bar h}_{45}^\prime
+\Phi_1{\bar\Phi}_1+\Phi_2{\bar\Phi}_2)\cr
&\qquad
+h_3{\bar h}_{45}{\bar\Phi}_{45}^\prime
+{\bar h}_3h_{45}{\Phi}_{45}^\prime
+h_3{\bar h}_{45}^\prime\Phi_{45}
+{\bar h}_3h_{45}^\prime{\bar\Phi}_{45}\cr
&\qquad
+{1\over2}(\xi_1D_1{\bar D}_1+\xi_2D_2{\bar D}_2)
+{1\over\sqrt{2}}(D_1{\bar D}_2\phi_2+{\bar D}_1D_2{\bar\phi}_1)\}
,\quad&(7)\cr}$$
where a common normalization constant ${\sqrt 2}g$ is assumed.

The ``anomalous" $U(1)_A$ is broken
by the Dine-Seiberg-Witten mechanism [\DSW]
in which a potentially large Fayet-Illiopolus D--term
is generated by
the VEV of the dilaton field. Such a D--term would, in general,
break supersymmetry and destabilize the string  vacuum,
unless there is a direction
in the scalar potential $\phi=\sum_i{\alpha_i\phi_i}$
 which is F flat
and also D flat with respect to the non anomalous gauge symmetries and in which
$\sum_i{Q_i^A{\vert\alpha_i\vert}^2}< 0$. If such a direction exists, it
will acquire a VEV, canceling the anomalous D--term, restoring supersymmetry
and stabilizing the vacuum. Since the fields
corresponding to such a flat
direction typically also carry charges for the non anomalous D--terms,
 a non trivial set of constraints
 on the possible choices of VEVs is imposed. A particular example, in the
model under consideration, is given by the set
$\{\Phi_{45},\Phi^\prime_{45}\}$ with
$\vert\langle\Phi_{45}\rangle\vert^2=
3\vert\langle\Phi^\prime_{45}\rangle\vert^2=
{{3g^2}\over{16\pi^2}}$.

\bigskip
\noindent{\bf 3. Gauge Coupling Unification}

In this section I show that superstring gauge coupling unification can be
realized in the standard--like model, provided that the additional color
triplets and electroweak doublets exist at the appropriate scales.
The one--loop gauge couplings beta functions are given by
$${{d\alpha_i}\over{dt}}={{\alpha_i^2}\over{2\pi}}b_i\eqno(8)$$
where $\alpha_i\equiv{{g_i^2}\over{4\pi}}$ and $t=ln\mu$ is the energy scale.

The beta function coefficients in a
${SU(3)}\times {SU(2)\times {U(1)_Y}}$ basis are given by
\parindent=-15pt
1. For the minimal supersymmetric standard model,
$$b_i={\left(\matrix{-9\cr
                     -6\cr
                      ~0\cr}\right)}+
      {\left(\matrix{2\cr
                     2\cr
                     2\cr}\right)}n_{_G}+
      {\left(\matrix{0\cr
                     {1\over2}\cr
                     {3\over{10}}\cr}\right)}n_h,\eqno(9a)$$
where $n_{_G}=3$ is the number of generations and $n_h=2$ is the
number of Higgs doublets.

2. For the additional color triplets and electroweak doublets
$$b_{D_1,{\bar D}_1,D_2,{\bar D}_2}={\left(\matrix{{1\over2}\cr
                      0\cr
                     {1\over5}\cr}\right)};
  b_{D_3,{\bar D}_3}={\left(\matrix{{1\over2}\cr
                      0\cr
                     {1\over{20}}\cr}\right)};
  b_{\ell,{\bar\ell}}={\left(\matrix{0\cr
                      {1\over2}\cr
                         0\cr}\right)}.\eqno(9b)$$
\parindent=15pt

The renormalization group equations are integrated from the Z mass $M_Z$
to the unification scale. In the presence of various mass thresholds,
$M_I$ ($I=1,\cdots,n$), the
one--loop gauge couplings are given by [\GHS]
$$\alpha_G^{-1}=\alpha_i^{-1}(M_Z)-{1\over{2\pi}}
\left[b_i\ln{M_G\over{M_Z}}+b_{i1}\ln{M_G\over{M_1}}+
b_{i2}\ln{M_G\over{M_2}}+\cdot\cdot\cdot\right]\eqno(9c)$$
where $b_{iI}$, $i=1,2,3$ are the contribution of the new thresholds
to the beta functions.

 I take the following values at the
Z scale
$$\eqalignno{\sin^2\theta_W&=0.2334\pm0.001,&(10a)\cr
              \alpha^{-1}&=127.8\pm0.2,&(10b)\cr
              \alpha_s&=0.115\pm0.015.&(10c)\cr}$$
Using the relations
$\sin^2\theta_W={{3\alpha_1}\over{3\alpha_1+5\alpha_2}}$,
$\alpha={{3\alpha_1\alpha_2}\over{3\alpha_1+5\alpha_2}}$ and
$\alpha_3(M_Z)=\alpha_s$, the initial values of the coupling constants
are obtained,
$$\eqalignno{\alpha^{-1}&=58.83\pm0.11,&(11a)\cr
             \alpha^{-2}&=29.85\pm0.11,&(11b)\cr
             \alpha^{-3}&=8.70\pm1.15.&(11c)\cr}$$
The unification scale is defined as the scale at which
the gauge couplings are equal
$$\alpha_3(M_U)=\alpha_2(M_U)=\alpha_1(M_U).\eqno(12)$$
In the minimal supersymmetric model a unification scale at
$O(10^{16}GeV)$ is consistent with LEP precision data.
In the superstring standard--like models the unification scale
is pushed to
$$M_U\sim1\times 10^{18}GeV{\hskip 2cm}\alpha_U=0.059,\eqno(13)$$
provided, for example,
that the exotic color triplets and electroweak doublets
masses are
$$M_{D_1,{\bar D}_1,D_2{\bar D}_2}
\sim O(10^{8}GeV); M_{D_3,{\bar D}_3}\sim O(10^{12}GeV)\eqno(14a)$$
and
$$M_{\ell,{\bar\ell}}\sim O(4\times10^{11}GeV).\eqno(14b)$$
Clearly other possibilities for the masses of the
additional states do exist,
this being just a particular example which realizes superstring
gauge coupling unification.

\noindent{\bf 4. Conclusions}

In this paper I have shown that
in a class of superstring derived standard--like
models, string gauge coupling unification
can be consistent with precision LEP data for $\sin^2\theta_W$
and $\alpha_s$ . I showed that the presence of exotic color triplets
and electroweak doublets in the massless spectrum of superstring
standard--like models can elevate the unification scale to
$O(10^{18})GeV$, provided that the additional states exist at the
appropriate mass scales. The particular model under consideration
has three additional color triplets, in vector--like representations,
beyond the minimal supersymmetric standard model. The number
of additional color triplets and electroweak doublets depends
on the boundary condition basis vectors and on the choice of
generalized GSO projection coefficients. For example, the GSO
projection coefficients, Eq. (1), results in the additional
color triplets $D_1,{\bar D}_1,D_2,{\bar D}_2$. In the model
of Ref. [\TOP], where the opposite sign for
$c\left(\matrix{\gamma\cr
                                    1\cr}\right)$ was taken,
we obtain electroweak doublets instead of color triplets.
Thus, the number of additional color triplets and electroweak
doublets is model dependent. Consequently, whether or not superstring
gauge coupling unification can be realized is highly model dependent.
However, we may conclude that superstring gauge coupling unification
in the standard--like models is possible.
The model of table 1, being an explicit example that contains all the necessary
massless states to achieve superstring gauge coupling unification.

\centerline{\bf Acknowledgments}
I thank Uri Sarid for useful discussions. This work is supported
in part by a Feinberg School Fellowship.

\baselineskip=12pt
\refout

\vfill
\eject

\input tables.tex
\magnification=1000
\baselineskip=18pt
\hbox
{\hfill
{\begintable
\ F \ \|\ SEC \ \|\ $SU(3)_C$ $\times$ $SU(2)_L$ \ \|\ $Q_C$ & $Q_L$ & $Q_1$ &
$Q_2$
 & $Q_3$ & $Q_4$ & $Q_5$ & $Q_6$ \ \|\ $SU(5)$ $\times$ $SU(3)$ \ \|\ $Q_7$ &
$Q_8$  \crthick
$D_1$ \|\ ${b_2+b_3+\beta}$ \|(3,1)\|~~$1\over4$ & ~~$1\over2$
 & ~~${1\over4}$ & $-{1\over 4}$ &
 ~~${1\over 4}$ & ~~0 & ~~0 & ~~0 \|(1,1)\| $-{1\over4}$ &
$-{{15}\over 4}$   \nr
${\bar D}_1$ \|\ ${\pm\gamma}+(I)$
\|(${\bar 3}$,1)\| $-{1\over4}$ &
$-{1\over2}$ & $-{1\over4}$ & ~~${1\over 4}$ &
 ~~${1\over 4}$ & ~~0 & ~~0 & ~~0
\|(1,1)\| ~~${1\over 4}$ & ~~${{15}\over 4}$  \cr
$D_2$ \|\ ${b_1+b_3+\alpha}$ \|(1,1)\|~~$1\over4$ & ~~$1\over2$
 & $-{1\over4}$ & ~~${1\over 4}$ &
 $-{1\over 4}$ & ~~0 & ~~0 & ~~0 \|(1,1)\| $-{1\over 4}$ &
$-{{15}\over 4}$   \nr
${\bar D}_2$ \|\ ${\pm\gamma}+(I)$
\|(${\bar 3}$,1)\| $-{1\over4}$ & $-{1\over2}$
& ~~${1\over4}$ & $-{1\over 4}$ &
 ~~${1\over 4}$ & ~~0 & ~~0 & ~~0
\|(1,1)\| ~~${1\over 4}$ & ~~${{15}\over 4}$  \cr
$D_3$ \|\ ${1+\alpha}$ \|(3,1)\|~~$1\over2$ & ~~0
 & ~~0 & ~~0 &
 ~~0 & ~~${1\over 2}$ & ~~${1\over2}$ &
{}~~${1\over2}$ \|(1,1)\| $-1$ &
{}~~0   \nr
${\bar D}_3$ \|\ $+2\gamma$
\|(${\bar 3}$,1)\| $-{1\over2}$ & ~~0 & ~~0 & ~~0 &
 ~~0 & $-{1\over 2}$ & $-{1\over 2}$ & ~~${1\over2}$
\|(1,1)\| ~~1 & ~~0  \cr
$\ell_1$ \|\ ${1+b_3+\alpha}$ \|(1,2)\| ~~0 & ~~0
 & ~~0 & ~~0 &
 ~~${1\over 2}$ & ~~${1\over 2}$ & ~~${1\over 2}$
& ~~0 \|(1,1)\| ~~1 &
{}~~0   \nr
${\bar\ell}_1$ \|\ $2\gamma$
\|(1,2)\| ~~0 & ~~0 & ~~0 & ~~0 &
 $-{1\over 2}$ & ~~${1\over 2}$ & ~~${1\over 2}$ & ~~0
\|(1,1)\| $-1$ & ~~0  \cr
$\ell_2$ \|\ ${1+b_2+\alpha}$ \|(1,2)\|~~0 & ~~0
 & ~~0 & ~~${1\over 2}$ &
 ~~0 & ~~${1\over 2}$ & ~~0 & ~~${1\over 2}$ \|(1,1)\| ~~1 &
{}~~0   \nr
${\bar\ell}_2$ \|\ $2\gamma$
\|(1,2)\| ~~0 & ~~0 & ~~0 & $-{1\over 2}$ &
 ~~0 & ~~${1\over 2}$ & ~~0 & ~~${1\over 2}$
\|(1,1)\| $-1$ & ~~0  \cr
$\ell_3$ \|\ $1+b_1+\alpha$ \|(1,2)\| ~~0 & ~~0
 & ~~${1\over2}$ & ~~0 &
 ~~0 & ~~0 & ~~${1\over2}$ & ~~${1\over2}$
\|(1,1)\| ~~1 &
{}~~0  \nr
${\bar\ell}_3$ \|\ $2\gamma$
\|(1,2)\| ~~0 & ~~0
 & $-{1\over2}$ & ~~0 &
 ~~0 & ~~0 & ~~${1\over2}$ & ~~${1\over2}$
\|(1,1)\| $-1$ &
{}~~0
\endtable}
\hfill}
\bigskip
\parindent=0pt
\hangindent=39pt\hangafter=1
{\it Table 2.} The additional color triplets and electroweak doublets,
in the massless spectrum of the model of table 1, and their quantum numbers.

\vfill
\eject

\input tables.tex

\hoffset=1.5truein
\magnification=1000
\font\normalroman=cmr10
\font\style=cmr7
\style

\fontdimen12\fivesy=0pt

\textfont0=\sevenrm
\scriptfont0=\fiverm
\textfont1=\seveni
\scriptfont1=\fivei
\textfont2=\sevensy
\scriptfont2=\fivesy

{\hfill
{\begintable
\  \ \|\ $\psi^\mu$ \ \|\ $\{\chi^1;\chi^2;\chi^3\}$ \ \|\
$y^3y^6$,  $y^4{\bar y}^4$, $y^5{\bar y}^5$,  ${\bar y}^3{\bar y}^6$
\ \|\ $y^1\omega^5$,  $y^2{\bar y}^2$,  $\omega^6{\bar\omega}^6$,
${\bar y}^1{\bar\omega}^5$
\ \|\ $\omega^2{\omega}^4$,  $\omega^1{\bar\omega}^1$,
$\omega^3{\bar\omega}^3$,  ${\bar\omega}^2{\bar\omega}^4$ \ \|\
${\bar\psi}^1$, ${\bar\psi}^2$, ${\bar\psi}^3$,
${\bar\psi}^4$, ${\bar\psi}^5$, ${\bar\eta}^1$,
${\bar\eta}^2$, ${\bar\eta}^3$ \ \|\
${\bar\phi}^1$, ${\bar\phi}^2$, ${\bar\phi}^3$, ${\bar\phi}^4$,
${\bar\phi}^5$, ${\bar\phi}^6$, ${\bar\phi}^7$, ${\bar\phi}^8$ \crthick
$\alpha$
\|\ 0 \|
$\{0,~0,~0\}$ \|
1, ~~~1, ~~~~1, ~~~~0 \|
1, ~~~1, ~~~~1, ~~~~0 \|
1, ~~~1, ~~~~1, ~~~~0 \|
1, ~~1, ~~1, ~~0, ~~0, ~~0, ~~0, ~~0 \|
1, ~~1, ~~1, ~~1, ~~0, ~~0, ~~0, ~~0 \nr
$\beta$
\|\ 0 \| $\{0,~0,~0\}$ \|
0, ~~~1, ~~~~0, ~~~~1 \|
0, ~~~1, ~~~~0, ~~~~1 \|
1, ~~~0, ~~~~0, ~~~~0 \|
1, ~~1, ~~1, ~~0, ~~0, ~~0, ~~0, ~~0 \|
1, ~~1, ~~1, ~~1, ~~0, ~~0, ~~0, ~~0 \nr
$\gamma$
\|\ 0 \|
$\{0,~0,~0\}$ \|
0, ~~~0, ~~~~1, ~~~~1 \|\
1, ~~~0, ~~~~0, ~~~~0 \|
0, ~~~1, ~~~~0, ~~~~1 \|
 ~~$1\over2$, ~~$1\over2$, ~~$1\over2$, ~~$1\over2$,
{}~~$1\over2$, ~~$1\over2$, ~~$1\over2$, ~~$1\over2$ \| $1\over2$, ~~0, ~~1,
{}~~1,
{}~~$1\over2$,
{}~~$1\over2$, ~~$1\over2$, ~~0 \endtable}
\hfill}
\smallskip
\parindent=0pt
\hangindent=39pt\hangafter=1
\normalroman

{{\it Table 1.} A three generations $SU(3)\times SU(2)\times U(1)^2$ model.
The choice of generalized GSO coeficients is:
$c\left(\matrix{b_j\cr
                                    b_i,S\cr}\right)=
c\left(\matrix{b_j\cr
                                    \alpha,\beta,\gamma\cr}\right)=
-c\left(\matrix{\alpha\cr
                                    1\cr}\right)=
-c\left(\matrix{\alpha\cr
                                    \beta\cr}\right)=
-c\left(\matrix{\beta\cr
                                    1\cr}\right)=
-c\left(\matrix{\gamma\cr
                                    1\cr}\right)=
-c\left(\matrix{\gamma\cr
                                   \alpha,\beta\cr}\right)=
-1$ (j=1,2,3), with the others specified by modular invariance and space--time
supersymmetry.}
\vskip 2cm

\bye